\documentclass[twocolumn]{aastex631}
\usepackage{amsmath}
\usepackage{bm}

\shorttitle{Field-Level Comparison of N-body Simulations}
\shortauthors{Bayer,  Villaescusa-Navarro, et al.}
\begin{document}

\title{Field-Level Comparison and Robustness Analysis of Cosmological N-body Simulations}

\correspondingauthor{Adrian E. Bayer}
\email{abayer@flatironinstitute.org}

\author[0000-0002-3568-3900]{Adrian E. Bayer}
\affiliation{Center for Computational Astrophysics, Flatiron Institute, 162 5th Avenue, New York, NY 10010, USA}
\affiliation{Department of Astrophysical Sciences, Princeton University, Peyton Hall, Princeton, NJ 08544, USA}

\author[0000-0002-4816-0455]{Francisco Villaescusa-Navarro}
\affiliation{Center for Computational Astrophysics, Flatiron Institute, 162 5th Avenue, New York, NY 10010, USA}
\affiliation{Department of Astrophysical Sciences, Princeton University, Peyton Hall, Princeton, NJ 08544, USA}

\author[0000-0002-0869-8760]{Sammy Sharief}
\affiliation{Ciela Institute, Montréal, Canada}
\affiliation{Mila - Quebec Artificial Intelligence Institute, Montréal, Canada}
\affiliation{Department of Physics, Université de Montréal, Montréal, Canada}

\author[0000-0001-7689-0933]{Romain Teyssier}
\affiliation{Department of Astrophysical Sciences, Princeton University, Peyton Hall, Princeton, NJ 08544, USA}

\author[0000-0002-9853-5673]{Lehman H. Garrison}
\affiliation{Scientific Computing Core, Flatiron Institute, 162 5th Avenue, New York, NY 10010, USA}

\author[0000-0003-3544-3939]{Laurence Perreault-Levasseur}
\affiliation{Ciela Institute, Montréal, Canada}
\affiliation{Mila - Quebec Artificial Intelligence Institute, Montréal, Canada}
\affiliation{Department of Physics, Université de Montréal, Montréal, Canada}
\affiliation{Center for Computational Astrophysics, Flatiron Institute, 162 5th Avenue, New York, NY 10010, USA}
\affiliation{Perimeter Institute for Theoretical Physics, Waterloo, Canada}
\affiliation{Trottier Space Institute, McGill University, Montréal, Canada}

\author[0000-0003-2630-9228]{Greg L. Bryan}
\affiliation{Department of Astronomy, Columbia University, 550 West 120th Street, New York, NY, 10027, USA}

\author[0000-0001-6134-8797]{Marco Gatti}
\affiliation{Kavli Institute for Cosmological Physics, University of Chicago, Chicago, IL 60637, USA}

\author[0000-0003-2630-9228]{Eli Visbal}
\affiliation{Department of Physics and Astronomy and Ritter Astrophysical Research Center, University of Toledo, 2801 W. Bancroft Street, Toledo, OH 43606, USA}

\begin{abstract}
We present the first field-level comparison of cosmological N-body simulations, considering various widely used codes: Abacus, CUBEP$^3$M, Enzo, Gadget, Gizmo, PKDGrav, and Ramses. Unlike previous comparisons focused on summary statistics, we conduct a comprehensive field-level analysis: evaluating statistical similarity, quantifying implications for cosmological parameter inference, and identifying the regimes in which simulations are consistent. 
We begin with a traditional comparison using the power spectrum, cross-correlation coefficient, and visual inspection of the matter field.
We follow this with a statistical out-of-distribution (OOD) analysis to quantify distributional differences between simulations, revealing insights not captured by the traditional metrics.
We then perform field-level simulation-based inference (SBI) using convolutional neural networks (CNNs), training on one simulation and testing on others, including a full hydrodynamic simulation for comparison. 
We identify several causes of OOD behavior and biased inference, finding that resolution effects, such as those arising from adaptive mesh refinement (AMR), have a significant impact.
Models trained on non-AMR simulations fail catastrophically when evaluated on AMR simulations, introducing larger biases than those from hydrodynamic effects. Differences in resolution, even when using the same N-body code, likewise lead to biased inference. We attribute these failures to a CNN's sensitivity to small-scale fluctuations, particularly in voids and filaments, and demonstrate that appropriate smoothing brings the simulations into statistical agreement. Our findings motivate the need for careful data filtering and the use of field-level OOD metrics, such as PQMass, to ensure robust inference. 
%However, we show that field-level OOD does not necessarily lead to biased inference, suggesting that such metrics may be overly conservative.
\end{abstract}

%\keywords{}

%\setcounter{tocdepth}{2}  % adjust depth if needed
%\tableofcontents

\section{Introduction} \label{sec:intro}

Ongoing and upcoming cosmological surveys will generate vast datasets, requiring accurate \textit{models} of nonlinear structure formation and advanced statistical techniques to robustly \textit{extract information} about cosmological parameters from small scales.
To \textit{model} cosmic structure in the nonlinear regime,
numerous N-body simulation have been developed, each utilizing different underlying N-body solvers and algorithmic techniques. To \textit{extract information} in the nonlinear regime, there has been much work on developing various novel summary statistics, as well as towards lossless extraction of information from the entirety of the field---so called \textit{field-level inference}.

To \textit{model} nonlinear cosmic structure,
N-body simulations use different underlying algorithms to solve for the gravitational forces between simulated particles, balancing accuracy, resolution, and efficiency. For a system of $N$ particles, direct pairwise force calculations scale as $\mathcal{O}(N^2)$, which becomes prohibitively expensive for large $N$, motivating the development of more efficient methods.
Tree codes (e.g., PKDGrav3 \citep{PKDGrav}) approximate distant forces using a hierarchical structure, reducing complexity to $\mathcal{O}(N\log N)$. TreePM codes (e.g., Gadget \citep{Gadget}, Gizmo \citep{Gizmo}) combine a Particle-Mesh (PM) method---which also scales as $\mathcal{O}(N\log N)$---for long-range forces with a tree algorithm for short-range interactions. P$^3$M codes (e.g., CUBEP$^3$M \citep{2013MNRAS.436..540H}) refine the PM method by adding direct particle-particle interactions for better small-scale accuracy than pure PM, but are less efficient than TreePM. Similarly, Abacus \citep{2021MNRAS.508..575G} uses a particle-particle method for short-range interactions and multipole expansions for long-range, but with no overlap between short and long range. AMR codes (e.g., Enzo \citep{Enzo}, Ramses \citep{Ramses}) dynamically refine a grid where needed, increasing resolution in dense regions, unlike non-AMR codes which use a fixed softening length over a fixed mesh. These differences influence computational cost, force accuracy, and small-scale structure resolution, leading to variations in results across simulations. 

To perform \textit{inference}, simulation-based inference (SBI) is an emerging tool to determine the cosmological parameters by training a neural network on cosmological simulations \cite[see e.g.][]{Paco_2021a, Hahn:2023udg, Ho:2024whi, Tucci:2023bag}. While SBI can be applied to any summary of the data, such as the power spectrum, here we focus on \textit{field-level inference} which seeks to extract all of the information from every pixel in the field \citep[see e.g.][]{Jasche_2013, Seljak_2017, Jasche:2018oym, Schmidt:2018bkr, Schmidt:2020viy, Nguyen:2020hxe, Jeffrey:2020xve, Kostic:2022vok, Bayer_2023_vel, Bayer:2023rmj, SimBIG:2023ywd, Sharma:2024pth, Nguyen:2024yth, Euclid:2024ris, Jia:2024wbr, Horowitz:2025kop, Golshan:2024lmr, Parker:2025mtg}.

A key consideration regarding SBI is that the results are only as good as the simulations the neural network is trained on.
If there are large discrepancies between simulation and reality, also known as model misspecification or distribution shifts, this would lead to biases in results when the model is applied to the real Universe. This is true not just for machine learning methods, but also for Gaussian likelihoods where covariances are computed using simulations. One example of model misspecification in cosmological SBI is in the context of hydrodynamic simulations, where \cite{CAMELS_public, CMD}, and various works thereafter, found that different models of astrophysics in different hydrodynamic simulations have a large model misspecification issue --- training a neural network on one type of simulation and testing on another simulation (never mind the real Universe) leads to biased results.

More broadly in astrophysics and machine learning, there has been much recent attention to the sensitivity of SBI methods to distribution shifts \citep{Filipp:2024yef, Swierc:2024gdu, Agarwal:2024uel} and adversarial attacks \citep{Horowitz:2022uxh}, with some attempts made to alleviate these issues using domain adaptation and generalization \citep{AnauMontel:2022ppb, masserano2023}. To understand the implications for cosmological N-body simulations, we quantify the distributional shifts between the outputs different cosmological simulations by performing a statistical out-of-distribution (OOD) analysis using PQMass \citep{PQMass2024}.

Furthermore, we perform a sensitivity analysis to understand how differences in the outputs of cosmological simulations ultimately affect the results of cosmological parameter inference.
Previous cosmological sensitivity analyses have been performed by
\citet{Modi:2023llw}, who studied the impact of halo and galaxy modeling in N-body simulations for both the power spectrum and bispectrum, and \citet{Shao:2022mzk}, who studied the robustness of halos in hydrodynamic simulations using graph neural networks.
Here, we extend the sensitivity analysis to field-level inference at the map level, where the data corresponds to every pixel in maps of the matter field. We use a convolutional neural network (CNN) to extract the information.
We perform our sensitivity analysis on the underlying N-body particle distribution, without considering the matter-halo connection, as, if the underlying N-body simulations suffer from robustness issues, this could imply problems for inference using any derived quantities, whether it be related to halos, galaxies, weak lensing, or another observable. This is particularly important given different cosmological survey collaborations have elected to use different N-body simulations, for example DESI makes heavy use of Abacus, while Euclid predominantly uses PKDGrav3.

This work has three key unique contributions:
\begin{itemize}
    \item We take a step back from hydrodynamic simulations by comparing their underlying N-body solvers. Different hydrodynamic codes make use of different N-body solvers under the hood, thus a part of the model misspecification reported amongst different hydrodynamic simulations by \cite{CMD}, and works thereafter, may be due to differences in their respective N-body solvers. Moreover, N-body codes without hydrodynamic effects underlie many cosmological analyses, for example for galaxy clustering and weak lensing, thus understanding the sensitivity of different N-body codes to model misspecification is imperative to ensuring accurate extraction of cosmological information. 
    \item While existing works comparing different N-body simulations consider a collection of standard summaries, such as the power spectrum and bispectrum \citep[e.g.][]{schneider, 2019MNRAS.485.3370G, Grove2022:DESIsim}, or use wavelets to quantify the noise in simulations \citep{Romeo:2008jx}, it is known that these do not fully unpack the total information in the cosmic field. Thus, here we compare the simulations at the field level. In particular, we perform a statistical OOD analysis using PQMass \citep{PQMass2024} to quantify the comparability of the density fields produced by different simulations.
    \item Crucially, we follow the analysis all the way through to investigate the impact on the inferred cosmological parameters by performing field-level simulation-based inference. This provides additional insight, as OOD at the data-vector level does not necessarily imply biased inference.
\end{itemize}

The paper is structured as follows. We discuss our method in Section \ref{sec:method}, present our results in Section \ref{sec:results}, and conclude in Section \ref{sec:conc}.

\section{Method} \label{sec:method}

We describe the simulations data used in \ref{sec:sim}, the OOD quantification procedure in \ref{sec:ood_method}, and the neural network architecture, training, and testing procedure in \ref{sec:nn}.

\subsection{Simulations}\label{sec:sim}

%\begin{centering}
\begin{table*}
\centering
	\setlength{\tabcolsep}{6pt}     
	\renewcommand{\arraystretch}{1}  
		\begin{tabular}{|c|c|c|c|}
			\hline
			Name & Method & LH Realizations & Reference \\
			\hline
			\hline
			Abacus & PP + multipole & 50 & \citet{2021MNRAS.508..575G} \\
			CUBEP$^3$M & P$^3$M & 50 & \citet{2013MNRAS.436..540H} \\
			Enzo & AMR & 0 & \citet{Enzo}\\
			Gadget & TreePM & 1000 & \citet{Gadget} \\
                Gizmo & TreePM & 0 & \citet{Gizmo} \\
			PKDGrav3 & Tree & 50 & \citet{PKDGrav} \\
			Ramses & AMR & 50 & \citet{Ramses} \\
			\hline 
                \hline
                IllustrisTNG (CAMELS) & Hydro & 50 & \citet{villaescusanavarro2020camels} \\
                \hline
		\end{tabular}
		%}
	\caption{\textbf{Characteristics of the simulations used in this work.} For each simulation we run a single matched seed realization as well as a Latin Hypercube set of varying size. We use 1000 Gadget realizations to train our neural network, and 50 realizations from each simulation for testing. We run no Latin Hypercube analysis for Enzo or Gizmo. We additionally ran a test using 50 full hydrodynamic simulations, namely  IllustrisTNG from the CAMELS suite.}
	\label{table:sims}
\end{table*}
%\end{centering}

We consider seven different N-body simulations, listed in Table \ref{table:sims}. All simulations follow the evolution of $256^3$ dark matter particles from $z=127$ down to $z=0$ in a volume of $(25\,{\rm Mpc}/h)^3$. The initial conditions have been generated at $z=127$ using second-order Lagrangian perturbation theory (except for the CUBEP$^3$M Latin Hypercube simulations, which used the Zel'dovich approximation). The $z=0$ particle snapshot of each simulation is painted and projected to a 2D image, with thickness $5\,{\rm Mpc}/h$, using a spherical kernel interpolation scheme (SPH), as done in \cite{CMD}. 

We run simulations with different seeds in a Latin Hypercube setup with parameter ranges $0.1 \leq \Omega_{\rm m} \leq 0.5$ and $0.6 \leq \sigma_8 \leq 1.0$, as well as 1 simulation with matched seed and cosmology ($\Omega_m=0.3175$ and $\sigma_8=0.834$). 
The other cosmological parameters are fixed at $\Omega_b=0.049$, $h=0.6711$, $n_s=0.9624$, $M_\nu=0$, and $w=-1$. 
For Gadget we generate 1000 simulation in the Latin Hypercube for training, while for the other codes we use $\sim 50$ simulations drawn from the same distribution of cosmological parameters for testing, except for Enzo and Gizmo for which we do no Latin Hypercube analysis.
For the matched seed datasets we ran an extra set of simulations for each code with double the particle resolution.
The Latin Hypercubes are used to quantify the precision and accuracy on parameter inference, while the
matched seed and cosmology simulations are used to understand the robustness in a controlled setting.

We now briefly describe each of the seven N-body simulations used:
\begin{enumerate}
\item \textbf{Gadget}. A TreePM code detailed in \citet{Gadget}. As for the CAMELS project, we use Gadget-III. The PM grid contains $1024^3$ voxels, and the FFT computations are performed using double precision. The following hyperparameter values were set: \texttt{ErrTolIntAccuracy}=0.025, \texttt{MaxSizeTimestep}=0.005, \texttt{ErrTolTheta}=0.5, \texttt{ErrTolForceAcc}=0.005, \texttt{TreeDomainUpdateFreq\\uency}=0.01. The softening length is set to 1/40 of the mean inter-particle distance.

\item \textbf{Abacus}. A particle-particle code described in \citet{2021MNRAS.508..575G}. The mathematical method is in \citet{2009PhDT.......175M}, and the softening scheme is validated in \citet{2021MNRAS.504.3550G}. Abacus uses an exact near-field/far-field force decomposition. The code uses particle pairs in the near field interact via direct pairwise evaluation, while the far-field forces employ a high-order multipole approximation. The resulting force errors are minimal, with a median fractional error of $\mathcal{O}(10^{-5})$. Leap frog integration is performed using a global time step, whose size is determined at each step. For enhanced accuracy, the time step parameter \texttt{TimeStepAccel} is set to $0.15$, as opposed to the usual $0.25$. The time step is relatively small compared to the dynamic times outside cluster cores, with the simulation taking about 2,000 steps to reach $z=0$. Spline softening is applied, with a softening length fixed in proper coordinates to a Plummer-equivalent length of $\ell/40$, capped at $0.3\ell$ in comoving coordinates. For the fixed-seed simulations used in Section \ref{sec:smooth}, the softening was fixed in comoving coordinates.

\item \textbf{CUBEP$^3$M}. A particle-particle particle-mesh (P$^3$M) code described in \citet{2013MNRAS.436..540H}. The gravitational force over large distances is computed using a two-level particle mesh method, while subgrid resolution is obtained via direct particle-particle interactions. High accuracy parameters for the force calculation, as tested in \citet{PhysRevD.100.083528}, were used. Additionally, we found that the code ran faster and produced better results when using a higher ratio of grid cells to particles (64:1 instead of 8:1) at a fixed softening length of $4.9 {\rm kpc}/h$. For the simulation with matched cosmology and random seed, we used the same initial particles as in the other codes.%, while the CUBEP$^3$M initial conditions, generated using the Zeldovich approximation, were used for latin-hypercube simulations.

\item \textbf{Enzo}. An Adaptive Mesh Refinement (AMR) code described in \citet{Enzo}, which uses a fast Fourier method \citep{Hockney_1998} to solve Poisson's equation on the root grid and a multigrid method on the subgrids. The dark matter particles are evolved using a kick-drift algorithm that provides second-order accuracy. Refinement occurs when the dark matter density in a cell exceeds $3\times 2^{3l}$ times the background density on the root grid, where $l$ is the refinement level. Our simulation includes 7 refinement levels. The parameters \texttt{CourantSafetyNumber} and \texttt{ParticleCourantSafetyNumber} were set to 0.15 and 0.125, respectively, to ensure smaller time steps and prevent large particle displacements relative to the most refined cell.

\item \textbf{Gizmo}. Developed as an extension of the Gadget code, Gizmo \citep{Gizmo} is a hybrid TreePM algorithm. Long-range forces are calculated using a particle-mesh (PM) method, which solves the Poisson equation in Fourier space to efficiently capture large-scale gravitational interactions. Short-range forces are computed using a hierarchical Barnes-Hut tree algorithm, where nearby particles are treated individually and distant groups of particles are approximated using a multipole expansion (up to the quadrupole level). An adaptive opening criterion ensures a balance between accuracy and computational efficiency. The gravitational forces are integrated using a symplectic kick-drift-kick (KDK) leapfrog scheme, ensuring good energy conservation over long timescales.
Gizmo is the underlying N-body code for the SIMBA hydrodynamical simulations.

\item \textbf{PKDGrav3}. Described in \citet{PKDGrav}, this code calculates particle forces using a highly efficient and memory-conserving version of the Fast Multipole Method (FMM, \citealt{Greengard1987}), with typical run-times scaling linearly with the number of particles. The FMM algorithm uses a binary tree structure, which reduces the number of terms in the multipole expansion required to evaluate the forces on particles. PKDGrav3 performance can be further enhanced by using GPU-accelerated nodes, which are used for particle-particle, particle-cell interactions, and periodic boundary conditions (via the Ewald summation method). Standard CPUs are used to build and traverse the trees.

\item \textbf{Ramses}. An AMR code based on the technique described in \citet{Ramses}. The AMR framework is based on a graded octree, where cells are refined individually when the enclosed mass exceeds a certain multiple $n_{\rm max}$ of the particle mass. For our initial analysis we refine when a cell has $n_{\rm max}=8$ times the particle mass, and also investigate the sensitivity to this parameter using $n_{\rm max}=2$ (denoted `Ramses2') for one specific part of our analysis in \ref{sec:smooth}. Mass deposition is done using the Cloud-In-Cell method. Poisson's equation is solved level by level using Dirichlet boundary conditions from the coarser level and a Multigrid relaxation solver. Time integration is performed with the Verlet algorithm (also known as adaptive leapfrog). The minimum refinement level was set to $\ell_{\rm min}=8$, corresponding to 256$^3$ base grid cells (one per particle on average), and the maximum level was $\ell_{\rm max}=15$, with a minimum cell size of about $1\,{\rm kpc}/h$. Note that the refinement strategy we have adopted here for the Enzo code corresponds to $n_{\rm max}=3$. 

\end{enumerate}

While the primary focus of our work is to consider N-body simulations, we additionally tested our network on a hydrodynamic simulation, \textbf{IllustrisTNG} from the CAMELS suite \citep{villaescusanavarro2020camels} to investigate the model misspecification when training on N-body and testing on hydrodynamic. These simulations were run with the AREPO code \citep{Arepo, Arepo_public}, for which the underling N-body solver is similar to Gadget, using TreePM plus moving-mesh finite volume (MMFV), and using the IllustrisTNG subgrid model \citep{WeinbergerR_16a,PillepichA_16a}.
We used 50 realizations with varying $\Omega_m$, $\sigma_8$, supernova, and AGN feedback parameters.

\subsection{Out-of-Distribution Analysis} \label{sec:ood_method}
We utilize PQMass \citep{PQMass2024} to compare the distribution of field-level outputs produced by different N-body simulations, as a means to quantify how OOD they are with respect to one another (using Gadget as a reference).
PQMass is a sample-based, likelihood-free framework designed to assess the performance of generative models by analyzing only their generated samples. Given two independent and identically distributed (i.i.d.) sets of samples, PQMass estimates whether these samples come from the same underlying distribution. To do so, the sample space is partitioned into measurable, non-overlapping regions using a Voronoi tessellation and the number of samples within each region is aggregated. Because PQMass only requires distance calculations between samples, it effectively scales to high-dimensional data, thus enabling a comprehensive full-field analysis without necessitating the dimensionality reduction commonly employed with standard summary statistics.

Given i.i.d.~samples in each set, the counts in each Voronoi region follow a multinomial distribution given by the underlying probability mass in each region. Under the null hypothesis that the two sets of samples come from the same underlying distribution, PQMass uses a $\chi^2$ test to compare the two sets of samples. Specifically, the PQMass test statistic, is given by
\begin{align}
\label{eqn:pqmass}
    \chi^2_{\rm {PQM}} := \sum_{j = 1}^{n_R} \Bigg[& \frac{\left(k\left({\bf x}_1, R_j\right) - \hat{N}_{j}^{(1)}\right)^2}{\hat{N}_{j}^{(1)}} \nonumber \\ 
    &+ \frac{\left(k\left({\bf x}_2, R_j\right) - \hat{N}_{j}^{(2)}\right)^2}{\hat{N}_{j}^{(2)}} \Bigg]\, ,
\end{align}
%\begin{align}
%\label{eq:chi2}
%    \chi^2_{\rm {PQM}} := \sum_{j = 1}^{n_R} \sum_{i=1}^{2} \frac{\left(k\left({\bf x}_i, R_j\right) - \hat{N}_{j}^{(i)}\right)^2}{\hat{N}_{j}^{(i)}},
%\end{align}
where $n_R$ is the number of regions, ${\bf x}_{i}$ is a vector containing samples from set $i\in\{1,2\}$,  $R_j$ is the underlying probability mass in region $j$, $k\left( {\bf x}_{i}, R_j\right)$ is the counts of samples from set $i$ in region $j$, and $\hat{N}_{j}^{(i)}$ is the expected number of samples from set $i$ in region $j$. 

The test statistic $\chi^2_{\rm PQM}$ follows a $\chi^2$ distribution with $n_R-1$ degrees of freedom. Any discrepancy with this quantifies how likely the difference in counts between the sets of samples is due to sample variance under the null hypothesis, and therefore how probable it is that the sets contain OOD samples. A larger deviation from the expected $\chi^2$ distribution implies a stronger degree of OOD.

In our analysis, we use the Latin Hypercube set of simulations for the samples in PQMass. Given the number of simulations available, we compare 50 simulations from each N-body solver to 500 simulations from Gadget. We partition the space into 50 regions and repeat the measurement 1000 times. 

\subsection{Neural Network for Parameter Inference}\label{sec:nn}

We use a CNN architecture which takes a $256^2$ input image and processes it through six convolutional blocks, each consisting of multiple convolutional layers, batch normalization, and LeakyReLU activations, progressively increasing the feature depth and downsampling spatial dimensions. The model uses circular padding to avoid edge artifacts. After flattening the output, two fully connected layers predict a 12-dimensional output (corresponding to the 6 parameter means and standard deviations), where the last six values are squared to enforce positivity. Dropout is applied for regularization. This architecture is optimized for learning hierarchical features from the input while minimizing overfitting. The architecture is described fully in the Appendix in Table \ref{tab:arch}.

We perform SBI to infer the cosmological parameters $\theta$ given the 2D density field data $X$ by using the loss
\begin{align} 
    \label{cmd_loss_fn}
    \mathcal{L} &= {\sum_{i=1}^{2}\log{\left(\sum_{j \in batch}(\theta_{i,j} - \mu_{i,j})^2\right)}} \nonumber \\
    &+ {\sum_{i=1}^{2}\log{\left(\sum_{j \in batch}((\theta_{i,j} - \mu_{i,j})^2 - \sigma_{i,j}^2)^2\right)}},
\end{align}
where $\theta_i$ is the true value of the $i^{\rm th}$ parameter. This loss predicts the mean and variance of the marginal posterior of the parameters,
\begin{equation}
    \mu_i({X}) = \int_{\theta_i}{p(\theta_i|\bm{X})\theta_id\theta_i},
\end{equation}
\begin{equation}
    \sigma_i({X}) = \int_{\theta_i}{p(\theta_i|\bm{X})(\theta_i - \mu_i)^2d\theta_i},
\end{equation}
where $p(\theta_i|\bm{X})$ is the marginal posterior over the $i^{th}$ parameter  \citep{moment_networks}. While other methods of SBI are available, we choose this method to mimic a common choice used for the CAMELS dataset \citep{CMD}.

To compare the field-level results of the network, and robustness to different N-body solvers, we train the network using Gadget simulations, and test using all of the different simulations. 
%The choice of Gadget is arbitrary, but motivated by the Quijote simulations, which is a large suite of simulations run with Gadget used for machine learning application \cite{quijote}.
The data is split into three different sets: training, validation, and testing in the ratio $90\%:5\%:5\%$, such that each 2D map comes from a different original simulations box.
During training, we perform data augmentation through rotations and flips to enforce invariance to these symmetries.
We use the Adam optimizer with a cyclic learning rate scheduler with minimum learning rate of $10^{-9}$. We use the \texttt{optuna} package \citep{Optuna} to perform hyperparameter optimization over the number of hidden layers $H$ (between 6 and 12), the dropout rate for the fully connected layers (between 0 and 0.9), the weight decay (between $10^{-8}$ and $10^{-1}$), and the maximum learning rate (between $10^{-5}$ and $10^{-3}$).
We consider various metrics of network performance, including the residual error $\epsilon$, the root mean square error (RMSE), the coefficient of determination $R^2$, and the $\chi^2$. 

%quantify the accuracy of the network using the Perason $R^2$ score where
%\begin{equation}
%R^2 = 1 - \frac{\sum_{i=1}^{n} (\theta_i - \mu_i)^2}{\sum_{i=1}^{n} (\theta_i - \bar{\theta})^2}.
%\end{equation}

\section{Results} \label{sec:results}

\begin{figure*}[ht!]
\includegraphics[width=\textwidth]{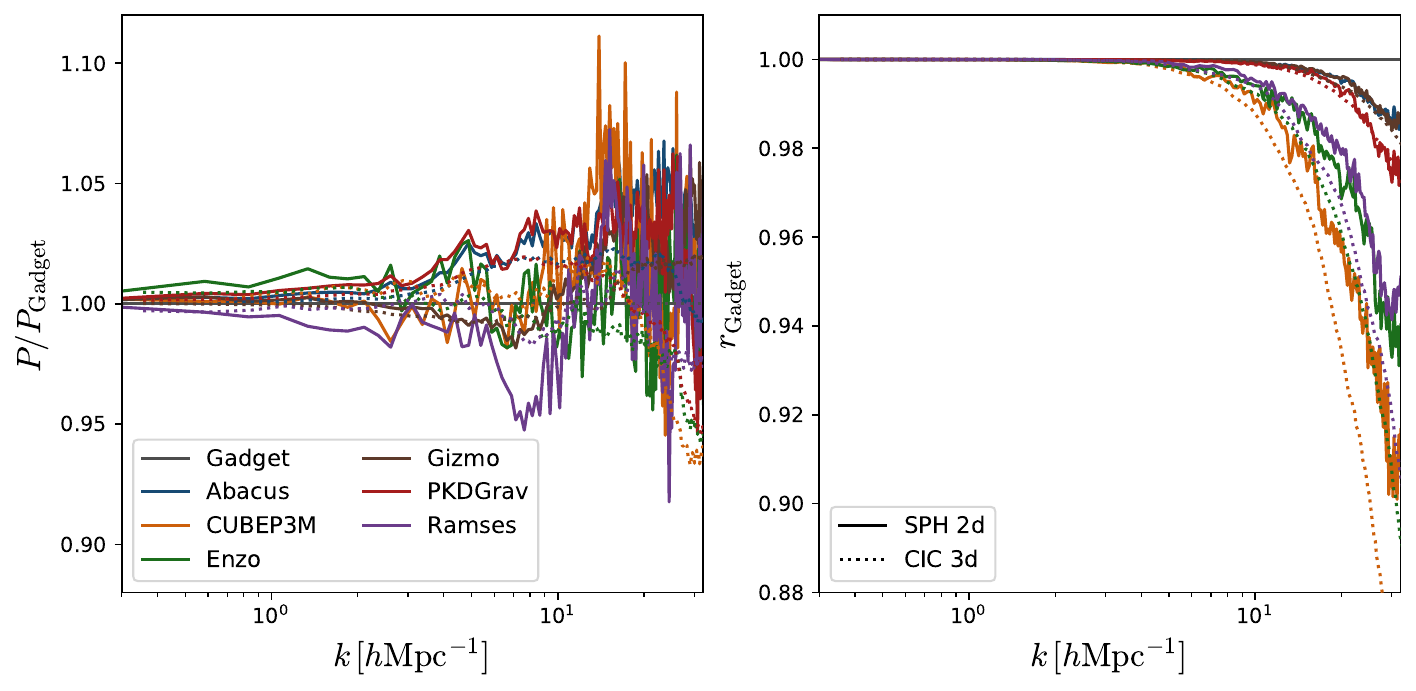}
\caption{
\textbf{The power spectrum (left) and cross correlation coefficient (right) of all N-body simulations with respect to Gadget with matched seed and cosmology.} We consider the power spectra of the 2D images, which were interpolated using an SPH-like kernel, and the power spectra of the 3D fields, which were interpolated using a CIC scheme. The power spectra all agree to within a few percent across all scales, with the two AMR codes (Enzo and Ramses) showing a larger disagreement with Gadget on large scales. The cross-correlation coefficient is lowest for CUBEP$^3$M, then the AMR simulations (Enzo and Ramses) are next lowest, with the non-AMR codes showing the best agreement.} 
\label{fig:power}
\end{figure*}

We first study the differences of the N-body simulations in terms of the power spectrum, cross-correlation, and images of the field in \ref{sec:power}. We then quantify how OOD the simulations are with respect to one another at the field level in \ref{sec:ood}. We then perform SBI to infer cosmological parameters in \ref{sec:lh}. We finally investigate the impacts of resolution and smoothing in \ref{sec:smooth}.

\subsection{Power, Cross-Power, and Field Comparison} \label{sec:power}

\begin{figure*}[ht!]
\includegraphics[width=\textwidth]{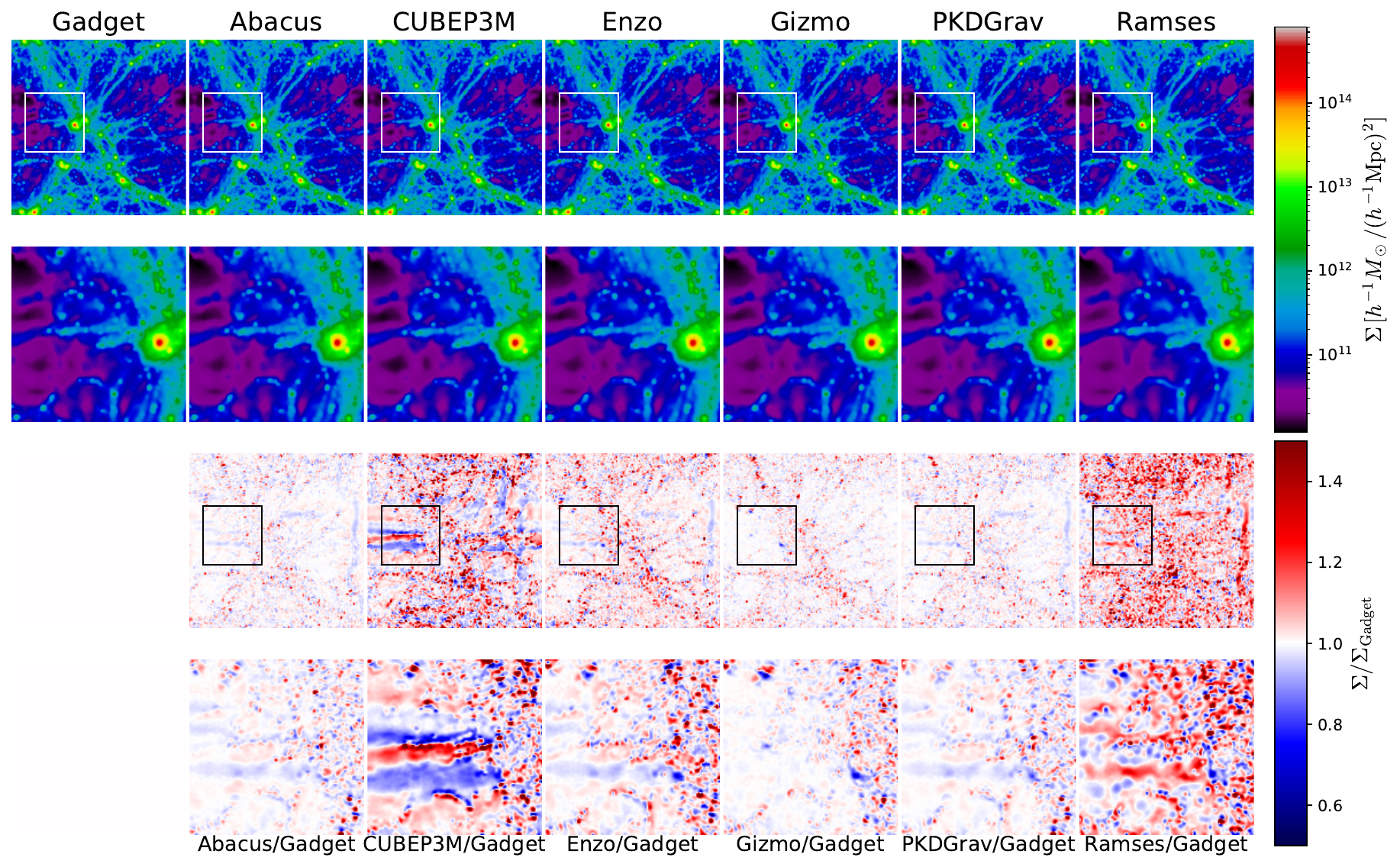}
\caption{\textbf{2D density fields $\Sigma$ produced by the different N-body codes with matched seed and cosmology.} The top row shows a $5\, {\rm Mpc}/h$ thick slice of the fields which look comparable by eye. The second row shows a zoom-in to the white box in the top row, where it can be seen that filaments and voids are smoother in the AMR codes, in particular for Ramses. The bottom two rows show the ratio of the density field of each simulation compared to Gadget, with and without zooming in, respectively. The non-AMR codes only show differences of $\lesssim 10\%$, apart from in very small structures in filaments and clusters where this increases to $\mathcal{O}(10\%)$. On the other hand, the AMR codes, particularly Ramses, show $\mathcal{O}(10\%)$ differences throughout the cosmic web, in particular along filaments and in voids. CUBEP$^3$M also shows significant disagreement throughout the cosmic web.
} 
\label{fig:slice}
\end{figure*}

We first compare the power spectra and cross-correlation coefficient of the different N-body simulations with the output of the Gadget simulation with a matched seed. The cross correlation coefficient between fields A and B is given by $r\equiv P_{AB}/\sqrt{P_AP_B}$ where $P_{AB}$ is the cross power of A and B, and $P_A$ and $P_B$ are the power spectra of A and B respectively. As the 2D maps used in this analysis are interpolated using an SPH method, following \cite{CMD}, we also show results for the 3D power spectrum and cross-correlation coefficient using CIC interpolation to show the overall trends are the same.

The left panel of Fig.~\ref{fig:power} shows the power spectra all agree to within a few percent across all scales. While all simulations become more discrepant on small scales, the two AMR codes (Enzo and Ramses) also show a small disagreement with Gadget on large scales \citep{Frenk:1999wm, 2019MNRAS.485.3370G}. Notably, CUBEP$^3$M maintains the best large-scale agreement with Gadget, up to $k\sim1\,h/{\rm Mpc}$.

The right panel of Fig.~\ref{fig:power} shows the cross-correlation coefficient between each simulation and the Gadget simulation. CUBEP$^3$M has the lowest agreement with Gadget. The AMR simulations (Enzo and Ramses) show the next lowest agreement with Gadget, with the non-AMR codes showing the best agreements. The cross-correlation coefficient quantifies the correlations in the phases of the field, thus, a decorrelation of order a few percent implies a considerable field-level difference between the different simulations, particularly between AMR and non-AMR simulations, and with CUBEP$^3$M.

To interpret the cause of this decorrelation, Fig.~\ref{fig:slice} shows images of the 2D density fields $\Sigma$ produced by the different N-body codes with matched seeds. The top row shows a $5\, {\rm Mpc}/h$ thick slice of the fields. The slices look somewhat comparable by eye. However, upon zooming in (second row), it can be seen that filaments and voids are smoother in the AMR codes, particularly for Ramses. For example, the lower left void in the zoom-in plot can be seen to be more underdense.

The bottom two rows of Fig.~\ref{fig:slice} show the ratio of the density field of each simulation compared to Gadget. The non-AMR codes only show differences of $\lesssim 10\%$, apart from in very small structures in filaments and clusters where this increases to $\mathcal{O}(10\%)$. While the eye is drawn to these strongly-speckled regions---indicating small-scale differences---this analysis does not distinguish differences in the overdense regions themselves from differences in the locations of overdense regions. In other words, some of the effect may be due to differences in large-scale evolution causing differences in overdensity location.

The AMR codes, particularly Ramses, show $\mathcal{O}(10\%)$ differences throughout the cosmic web, in particular along filaments and in voids. CUBEP$^3$M also shows significant disagreement throughout the cosmic web. 
One possible explanation for the large difference between CUBEP$^3$M and Gadget (even though both are non-AMR codes) is that it is due to resolution versus timing differences --- CUBEP$^3$M has the same resolution as Gadget, and so produces crisp small-scale structures, but has different timing for collapse of large-scale structure. It is also possible that the difference is caused by aliasing due to CUBEP$^3$M's use of Nearest Grid Point (NGP) interpolation for one of the PM force calculations --- CUBEP$^3$M mitigates the aliasing by adding a uniform offset to each particle coordinate at each time step, which could result in an overall offset in the positions of particles. Both of these effects could result in CUBEP$^3$M producing the same overall structures as Gadget, but at slightly different locations, which causes the decorrelation seen in Fig.~\ref{fig:power} and can be seen as small differences in the location of the large filament (big blue/red bands) in Fig.~\ref{fig:slice}.

\subsection{Field-Level Out-of-Distribution Analysis} \label{sec:ood}

\begin{figure*}[t]
\includegraphics[width=\textwidth]{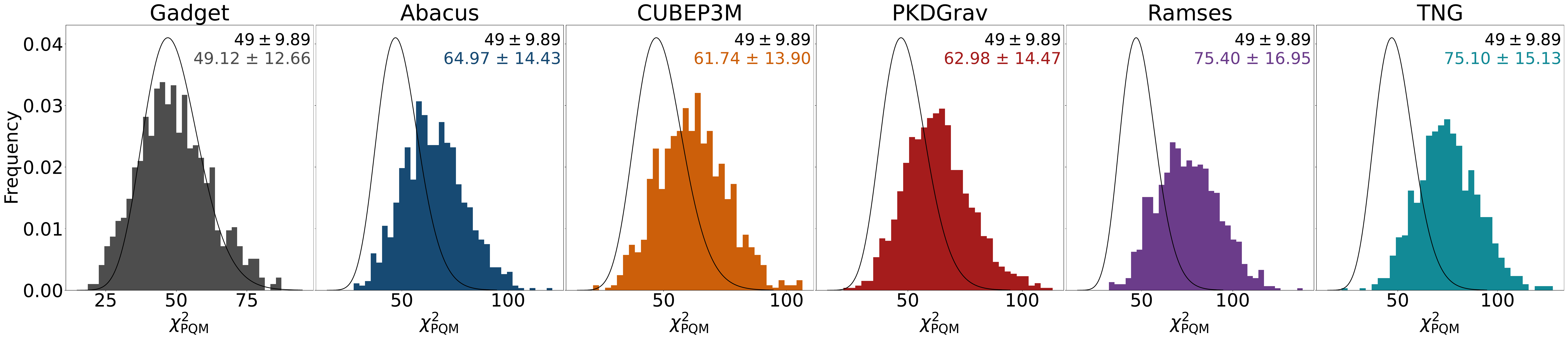}
\caption{\textbf{Field-level OOD analysis.} PQMass $\chi^2_{\rm PQM}$ distributions, comparing each simulation with respect to Gadget. Under the null hypothesis, and given our use of 50 comparison samples, when two sets of samples are from the same underlying distribution, $\chi^2_{\rm PQM}$ should follow a $\chi^2$ distribution with 49 degrees of freedom (black line). When comparing Gadget to Gadget, as a null test, $\chi^2_{\rm PQM}$ closely follows the $\chi^2$ distribution, implying Gadget is in distribution with itself. However, all other simulations tested against Gadget yield $\chi^2_{\rm PQM}$ distributions that deviate significantly from the expected $\chi^2$ behavior, indicating they are OOD relative to Gadget. Abacus, CUBEP$^3$M, and PKDGrav are similarly OOD with a mean $\chi^2_{\rm PQM}$ of between 61-65, while Ramses and TNG are the most OOD with a $\chi^2_{\rm PQM}$ of around 75.} 
\label{fig:PQMass}
\end{figure*}

\begin{figure*}[ht!]
\includegraphics[width=\textwidth]{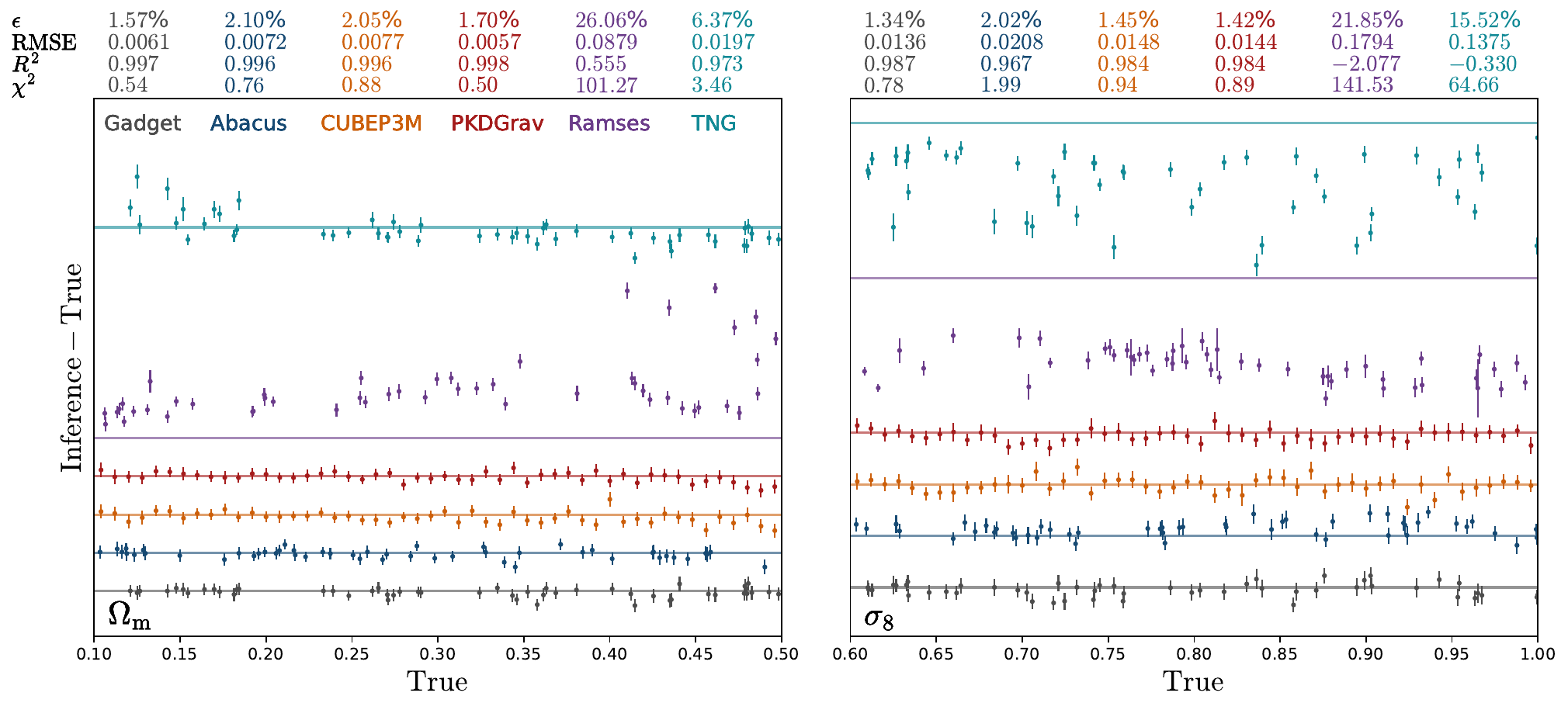}
\caption{
\textbf{Field-level SBI results when training on the Gadget Latin Hypercube and testing on other simulations} for $\Omega_m$ (left) and $\sigma_8$ (right). The true value of the parameters is denoted by the solid horizontal lines.
We did not run a Latin Hypercube for Enzo or Gizmo, but did additionally test on a hydrodynamic simulation (IllustrisTNG).
The error $\epsilon$, root mean square error (RMSE), coefficient of determination $R^2$, and $\chi^2$ are all reported on the top of the figure. 
Testing on other non-AMR codes (Gadget, Abacus, CUBEP$^3$M, PKDGRAV3) performs well, with $R^2\sim1$. Interestingly, even though CUBEP$^3$M shows a large disagreement in terms of the power spectrum, cross correlation, and image level comparison, the parameter constraints are not biased as a result. Meanwhile, testing on the AMR code (Ramses) there is a consistent large bias in the inference, namely a larger than $20\%$ error for both $\Omega_m$ and $\sigma_8$. 
When testing on the hydrodynamic simulation (TNG), the network is able to correctly predict $\Omega_m$ for some realizations, but is significantly biased for other realizations; it is also always biased for $\sigma_8$.
Note the spread of true values are different for the different codes due to the different selection of simulations and generation of the Latin Hypercubes.
Also note $\chi^2<1$ even for Gadget because the error bars inferred by the neural network are slightly overpredicted --- this could be improved by trained for longer or ensembling over many models, but this is unnecessary for the overall conclusions of this work.} 
\label{fig:inference_LH}
\end{figure*}

We now statistically analyze the distribution of simulations from a full field perspective using PQMass. For this analysis, we use all N-body simulations available in the Latin Hypercube suite: Gadget, Abacus, CUBEP$^3$M, PKDGrav, and Ramses. We also consider a comparison with TNG, a full hydrodynamic simulation, to quantify the distribution shift between a hydrodynamic simulation and a gravity-only N-body simulation.

Fig.~\ref{fig:PQMass} presents the distribution of $\chi^2_{\rm PQM}$ (eqn.~(\ref{eqn:pqmass})) obtained by comparing each simulation with respect to Gadget. As we use $n_R=50$ regions, the expected distribution is a $\chi^2$ distribution with 49 degrees of freedom, that is, a mean of 49 and a standard deviation of 9.89, depicted by the black line. As expected, the $\chi^2_{\rm PQM}$ from the Gadget null test is in agreement with the theoretical $\chi^2$ distribution, confirming internal consistency among Gadget samples (the small discrepancy in the standard deviation could be improved by using more samples). In contrast, all other simulations tested against Gadget yield $\chi^2_{\rm PQM}$ distributions that deviate significantly from the expected $\chi^2$ behavior, indicating that they are OOD relative to Gadget. We see that Abacus, CUBEP$^3$M, and PKDGrav are similarly OOD with a mean $\chi^2_{\rm PQM}$ of between 61-65, while Ramses and TNG are the most OOD with a $\chi^2_{\rm PQM}$ of around 75.

These results are mostly consistent with the power spectrum and cross correlation comparison in Section \ref{sec:power}, however, there is one key difference: while CUBEP$^3$M has the worst cross-correlation coefficient with respect to Gadget, it is similarly in distribution with Gadget as Abacus and PKDGrav. This implies that the cross-correlation coefficient and OOD detection metric provide complementary information---a cross-correlation coefficient and transfer function closer to unity does not guarantee a lower degree of field-level OOD.
This also lends credibility to the explanation given in Section \ref{sec:power}, that CUBEP$^3$M produces the same overall structures as Gadget but at slightly shifted locations.

%We emphasize that PQMass is not a physically motivated summary statistic but a statistical framework applied to the full-field data. It assesses the distributional similarity between sets of samples from the different simulations without relying on any dimensionality reduction or astrophysical observables. Thus, PQMass offers a complementary perspective to the inference analysis done in Section~\ref{sec:lh}. Although the PQMass analysis and SBI approach operate in fundamentally different spaces, the full-field density distribution versus a latent, cosmology-informative manifold, they both reveal the same conclusion: TNG, Ramses, Abacus, CUBEP$^3$M, and PKDGrav are all OOD compared to Gadget. 

\subsection{Field-Level Inference} \label{sec:lh}

\begin{figure*}[ht!]
\vspace{1em}
\includegraphics[width=\textwidth]{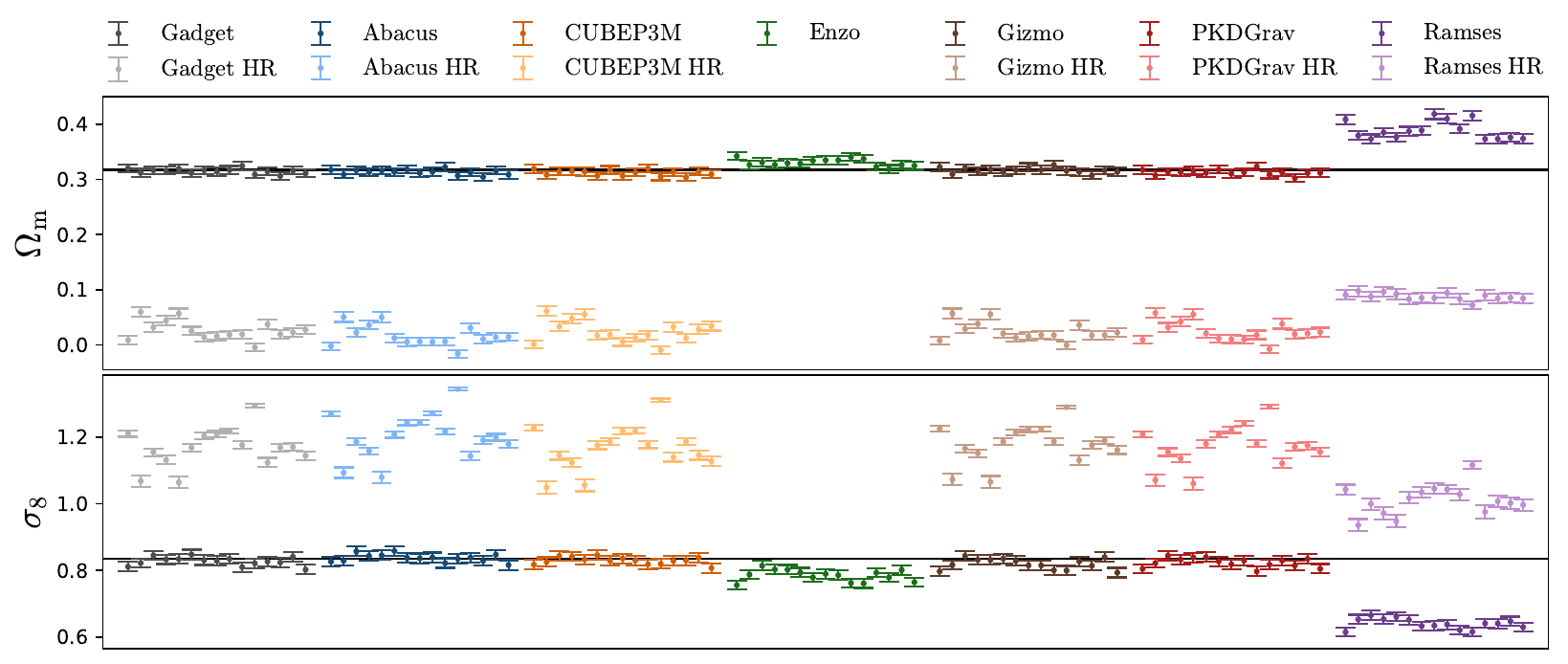}
\caption{
\textbf{SBI results when training on Gadget simulations, and testing on all simulations with fixed seed and cosmology} for $\Omega_m$ (top) and $\sigma_8$ (bottom). The 15 data points for each test correspond to 5 $5\,{\rm Mpc}/h$ slices that fit in the box sliced in all 3 directions. We perform tests at two particle resolutions, one where the network is tested at the same resolution as the training set, and another where the network is tested on simulations with twice the resolution (HR). When trained and tested at the same resolution, the agreement between non-AMR codes is overall good, while the AMR codes (Enzo and Ramses) show a bias, predicting a high $\Omega_m$ and a low $\sigma_8$. However, when testing the network on HR simulations we find a significant bias in all cases. This suggests that the model misspecification in AMR codes is due to differences in the effective resolution of the training and testing set, which particularly affects the AMR simulations as they have an adaptive resolution.} 
\label{fig:inference_seed}
\end{figure*}

Having compared the outputs of the different simulations, we now perform cosmological parameter inference. We train a CNN using the Latin Hypercube dataset of the Gadget simulations to infer the cosmological parameters $\Omega_m$ and $\sigma_8$. We then test this network on all of the different N-body simulations. Fig.~\ref{fig:inference_LH} shows the accuracy and precision of the inference for each simulation code for $\Omega_m$ (left) and $\sigma_8$ (right). The error $\epsilon$, root mean square error (RMSE), coefficient of determination $R^2$, and $\chi^2$ are all reported on the top of the figure. It can be seen that the model applied to other non-AMR codes (Gadget, Abacus, CUBEP$^3$M, PKDGRAV3) performs well as $R^2\sim1$. Interestingly, there is a slightly larger error when testing on Abacus for $\sigma_8$. It is also interesting to note that even though CUBEP$^3$M shows a large disagreement in terms of the cross-correlation and image level comparison, the parameter constraints are not biased as a result: this agrees with the OOD analysis in Section \ref{sec:ood} and is likely because CUBEP$^3$M produces the same structures as Gadget but at slightly shifted locations. Meanwhile, for the AMR code, Ramses, there is a large bias in the inference, namely a larger than $20\%$ error for both $\Omega_m$ and $\sigma_8$. 

To investigate the model misspecification when training on an N-body simulation and testing on a hydrodynamic simulation,
Fig.~\ref{fig:inference_LH} also shows results when the model trained on Gadget is tested on IllustrisTNG from the CAMELS suite \citep{villaescusanavarro2020camels}. In this case, the network is biased with respect to $\sigma_8$, which is expected due to the degeneracy with hydrodynamic effects \citep{Amon:2022azi}. We find that for many realizations the network is able to correctly predict $\Omega_m$, while for others---typically near the prior boundaries---it is significantly biased. It is interesting that the bias in cosmological parameters when testing on a hydrodynamic simulation is no worse than when testing on N-body Ramses. The results of IllustrisTNG is similar to Ramses in the sense that both infer a $\sigma_8$ that is biased low, however, for Ramses the bias is somewhat constant, while for IllustrisTNG it varies due to variations in baryonic parameters. Moreover, the IllustrisTNG results are different to Ramses in the sense that Ramses infers an $\Omega_m$ that is always biased high, while IllustrisTNG does not.

The results of this field-level parameter inference agree with the OOD analysis in Section \ref{sec:ood} in the sense that Ramses and TNG---which are the most OOD---also produce the most biased parameter constraints. However, while Abacus, CUBEP$^3$M, and PKDGrav were deemed OOD, they produce almost unbiased parameter constraints. This is not surprising, as even though two fields might be OOD, parameter inference essentially compresses a field into a summary of the total information which need not also be OOD.
This implies that a PQMass OOD analysis provides a useful diagnostic to determine whether parameter inference will be biased, but that it may also lead to overly conservative conclusions in terms of unbiased inference.

\subsection{Resolution is the Resolution} 
\label{sec:smooth}

\begin{figure*}[t]
\includegraphics[width=\textwidth]{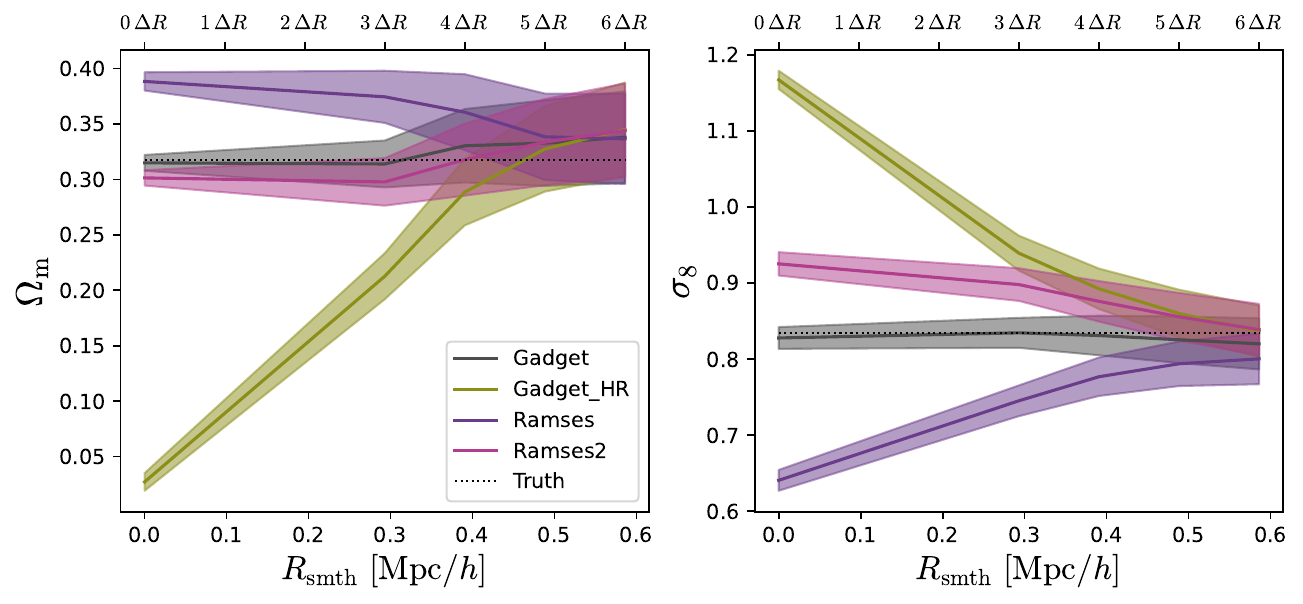}
\caption{
\textbf{Smoothing analysis.} SBI results when training on fixed seed and cosmology Gadget simulations, and testing on Gadget, high-resolution Gadget (Gadget\_HR), Ramses, and an excessively refined version of Ramses to bring it closer to a non-AMR simulation (Ramses2). The true value of the parameters is denoted by the dotted solid line. Before training and testing we smooth the image with a Gaussian filter with standard deviation given by integer multiples of the grid spacing $\Delta R \approx 0.1\,{\rm Mpc}/h$. It can seen that smoothing of around $\gtrsim 5 \Delta R$ is required for unbiased inference when the testing set does not match the training set. Also, while increasing the refinement of Ramses does improve the agreement without smoothing, some smoothing is still required for robust inference.} %Ramses: $m_refine=10*8$. Ramses2: $m_refine=0,2,2,2,2,2,2,2,2$. CAMELS Ramses: $m_refine=64,64,64,64$.} 
\label{fig:inference_smooth}
\end{figure*}

\begin{figure*}[ht!]
\includegraphics[width=\textwidth]{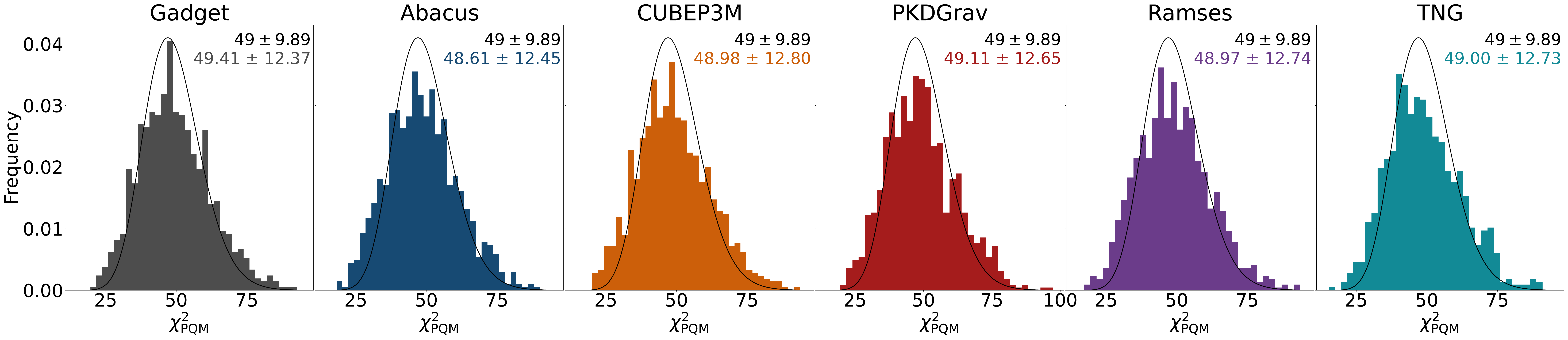}
\caption{
\textbf{Field-level OOD analysis on smoothed fields.} PQMass $\chi^2_{\rm PQM}$ distributions comparing each simulation to Gadget, after applying Gaussian smoothing with standard deviation $6 \Delta R$ ($\Delta R \approx 0.1\,{\rm Mpc}/h$). Under the null hypothesis that both sets of samples come from the same underlying distribution, $\chi^2_{\rm PQM}$ follows a $\chi^2$ distribution with 49 degrees of freedom (black line). After smoothing, all simulations, including Ramses and TNG, yield $\chi^2_{\rm PQM}$ distributions that closely match the expected behavior under the null. This is to be compared with Fig.~\ref{fig:PQMass} where no smoothing was applied.
}
\label{fig:PQMass_Smoothing_Analysis}
\end{figure*}

We now investigate the cause of the biased inference and large OOD, which implied a particularly strong difference between AMR and non-AMR solvers.
The key methodological difference in an AMR solver is the change in effective grid resolution due to the adaptive refinement, which suggests that the grid resolution may be the resolution. 
AMR codes will typically not adaptively refine the mesh inside low density regions, such as voids, leading to a smoother density compared to non-AMR codes which use a fixed softening length. On the other hand, non-AMR codes will find small structures inside voids, which may not be physical.
One can thus think of AMR codes as smoothing over anything that cannot be accurately resolved by the chosen number of simulation particles. 
This effective smoothing is particularly important for CNNs, as CNNs are sensitive to small scale fluctuations and thus could learn a lot of information from low density pixels (for example they might learn $\Omega_m$ from the particle mass or minimum halo mass if the grid is sufficiently sparse).

To investigate the effects of resolution we test our model from the previous section on simulations with the same seed and cosmology, at both the same resolution and also two times higher resolution (HR). 
While we did not have Latin Hypercube data for Enzo and Gizmo, we now test our inference on them too.
Fig.~\ref{fig:inference_seed} shows the inference of $\Omega_m$ (top) and $\sigma_8$ (bottom). When trained and tested at the same resolution, the agreement between non-AMR codes is overall good, while the AMR codes (Enzo and Ramses) show a bias, predicting a high $\Omega_m$ and a low $\sigma_8$.
When testing the network on maps from simulations with two times higher particle resolution (HR) we find that there is a significant bias in cosmological parameters in all cases. This is because higher resolution simulations find more small clusters, which the neural network has not been trained to notice. This suggests that the model misspecification in AMR codes is due to differences in their effective resolution with respect to fixed-grid simulations.

If it is the difference in small-scale structures caused by different resolutions that leads to this model misspecification, smoothing the fields should erase these differences and allow robust inference. We thus now test the effects of smoothing.
Fig.~\ref{fig:inference_smooth} shows the inferred cosmological parameters when training the model on a smoothed version of Gadget, and testing on a simulation with the same smoothing. To smooth, we apply a top-hat filter with size given by integer multiples of the grid spacing $\Delta R \approx 0.1\,{\rm Mpc}/h$. It can firstly be seen that testing on Gadget gives unbiased inference for all smoothing scales, which is expected as there is no model misspecification in this case. Secondly, it can be seen that when testing the model on the HR version of Gadget, there is severe model misspecification without smoothing, but when smoothing by $\gtrsim 6 \Delta R$, the inference is unbiased. Similarly, testing on Ramses gives biased results without smoothing, but smoothing by $\gtrsim 6 \Delta R$ results in unbiased inference. Note that this reduction in bias comes at the expense of less constraining power, signified by the wider error bands.

We additionally explore what happens when changing the mesh refinement used by Ramses: while our fiducial Ramses run refined any cell with $n_{\rm max}=8$ particles, we ran another set of Ramses simulations where any cell with $n_{\rm max}=2$ particles is refined (denoted `Ramses2'). A refinement criterion of 8 is seen as the minimum limit to enable a fluid-like simulation given the number of particles is $256^3$ which corresponds to on average one particle per grid cell. Using a higher level of refinement with $n_{\rm max}=2$ somewhat mimics a non-AMR code, as there will now be more mesh elements with very low (or even zero) particle counts. Fig.~\ref{fig:inference_smooth} shows that testing on Ramses2 is unbiased without smoothing for $\Omega_m$, but that $\sigma_8$ is now overpredicted without smoothing, albeit with a smaller bias. This implies that while arbitrarily refining the mesh in AMR simulations can somewhat alleviate the model misspecification issue, it alone cannot fully enable robust inference without smoothing.

Finally, we apply PQMass on maps smoothed by $6 \Delta R$.  Fig.~\ref{fig:PQMass_Smoothing_Analysis} shows the $\chi^2_{\rm PQM}$ distributions for all simulations now closely follow the expected $\chi^2$ distribution, indicating that smoothing brings everything in distribution. This reinforces the interpretation that differences in effective resolution and in small scale physics drive the observed biased inference via distribution shifts. This motivates computing PQMass as a function of smoothing scale as a practical approach to find the optimal smoothing scale for robust inference.

\section{Conclusions} \label{sec:conc}

We performed a \textit{field-level comparison and robustness analysis} of various commonly used cosmological N-body simulations: Gadget, Abacus, CUBEP$^3$M, Enzo, Gizmo, PKDGrav3, and Ramses.
We studied how the simulations appear different at the field level, finding a notable difference between simulations that do and do not use AMR to solve the N-body dynamics. We performed a \textit{field-level OOD analysis} using PQMass, showing this to be an informative diagnostic to determine the difference between two simulation suites at the field level.

Additionally, we  performed \textit{field-level simulation-based inference} of cosmological parameters with a CNN, training on Gadget (a non-AMR simulation) and testing on all other simulations. We found reasonable agreement when testing on any other non-AMR simulation, however, testing on AMR simulations can perform catastrophically---producing more biased results than when testing on a full hydrodynamic simulation. We also found that evaluating the model on the same N-body simulation run with a different resolution leads to biased inference. We attributed this to the CNN's sensitivity to small-scale fluctuations in voids and filaments caused by the simulation spatial resolution, advocating for careful smoothing to ensure robust inference. Moreover, while adjusting the refinement strategy of the AMR codes can somewhat alleviate the discrepancy, it alone cannot fully enable robust inference, and it is also not physically justified---one should sufficiently smooth over the scales that the N-body simulation, whether AMR or non-AMR, is unable to accurately resolve. In particular, we found Gaussian smoothing with a scale of $\sim6$ times the particle grid spacing is required for robust inference of $\Omega_m$ and $\sigma_8$, and to bring the simulations in distribution with one another. 

% Perhaps some discussion on the pros and cons of AMR vs non-AMR? Softening length etc.

% A discrepancy of only a few percent in the small scale cross correlation coefficient can cause catastrophic bias. A problem for tranfer function approaches

%Additionally, while parameter constraints of $\Omega_m$ and $\sigma_8$ are of key cosmological interest, N-body simulations are used for a great multitude of different applications. We thus performed a statistical out-of-distribution (OOD) analysis at the field level, finding two density fields being OOD with respect to one another is not informative of how OOD the parameter constraints are. For example, while AMR codes performed catastrophically badly in terms of parameter inference, they are not the most OOD in terms of their fields. This implies that OOD methods at the data-vector level may not be a good metric to determine whether the inference will be OOD. This makes intuitive sense, as it could be that one tiny change in the field can make the CNN get one specific feature of the field completely wrong, even though all other features of the field remain in distribution. Vice versa, a large change in parts of the field could still preserve certain features of the field, for example there may be conserved quantities. Alternatively, it could imply that at the dimensionality considered here ($256^2$) OOD methods do not fully quantify the level of OOD, or require a much larger training set. Further work along the lines of OOD quantification would be fruitful future work. 

We found that simulations that gave biased field-level parameter inference were also the most OOD, thus motivating OOD detection as a useful diagnostic to ensure robust inference---importantly finding that the cross-correlation coefficient and transfer function do not fully quantify the degree of field-level OOD. However, we also found that a field-level OOD test may lead to overly conservative conclusions, as even though two simulation might be OOD at the field level, parameter inference could still be unbiased. It would be fruitful future work to investigate this further, for example by defining thresholds for how large the OOD metric has to be to trigger an OOD detection, although this will likely be dependent on the downstream inference task.

We showed that smoothing the fields enabled robust inference between simulations with different resolutions. 
While techniques such as domain adaptation seek a shared latent space across simulations, here we instead use physically motivated filtering to align the simulations by removing scales where differences are expected.
It would be interesting future work to explore alternative methods of filtering the data to potentially more optimally ensure robust inference.
For example, \cite{arnab} showed that applying cuts in the density to exclude the most extreme regions does not remove much cosmological information regarding $\Omega_m$ and $\sigma_8$. 
It would also be fruitful to explore methods to perform robust inference by training using multiple datasets, such as \cite{Jo:2025ndh}.
Moreover, while \cite{Shao:2022mzk} considered graph neural networks applied to halos from these simulations, it would be interesting to apply the CNN approach used here on maps of halos or galaxies to find the cuts on parameters (such as minimum halo mass) required to ensure robustness between different N-body simulations and halo finders.

Our work highlights the importance of knowing the regimes in which simulations can be trusted when performing simulation-based field-level inference. While we focused on inferring $\Omega_m$ and $\sigma_8$ in $25\,{\rm Mpc}/h$ boxes, our findings apply broadly to any analysis that uses N-body simulations---even  in the absence of hydrodynamic effects---including galaxy surveys, weak lensing, and 21cm.

%% Also note that the akcnowlodgment environment does not support long amounts of text. If you have a lot of people and institutions to acknowledge, do not use this command. Instead, create a new 
\section*{Acknowledgments}
%\begin{acknowledgments}
We thank Raul Angulo, Derek Inman, and Mihir Kulkarni for useful discussions at the early stages of this work. The work of AEB, FVN, and LHG is supported by the Simons Foundation. SS and LPL acknowledge support from the Simons Collaboration on ``Learning the Universe'', the Canada Research Chairs Program, the National Sciences and Engineering Council of Canada through grants RGPIN-2020-05073 and 05102. The computations reported in this paper were performed using resources made available by the Flatiron Institute. The work is in part supported by computational resources provided by Calcul Quebec and the Digital Research Alliance of Canada. The analysis of the simulations has made use of the \textit{Pylians} library, publicly available at \url{https://github.com/franciscovillaescusa/Pylians3}.\\
%\end{acknowledgments}


%\vspace{5mm}
%\facilities{HST(STIS), Swift(XRT and UVOT), AAVSO, CTIO:1.3m,CTIO:1.5m,CXO}


%\software{astropy \citep{2013A&A...558A..33A,2018AJ....156..123A},  
  %        Cloudy \citep{2013RMxAA..49..137F}, 
     %     Source Extractor \citep{1996A&AS..117..393B}
        %  }


\bibliography{ref}{}
\bibliographystyle{aasjournal}



\appendix
\vspace{-3em}
%\section{CNN Architecture}
\begin{table*}[h]
\centering
\begin{tabular}{|p{8cm}|}
\hline
%\textbf{Detailed Architecture of \texttt{model\_o3\_err}} \\
%\hline
\textbf{Input:} Image of shape $(\texttt{channels}, 256, 256)$ \\ \hline

\textbf{Block 0:} \\
Conv2d(channels $\rightarrow$ 2H, kernel=3, stride=1, pad=1) \\
Conv2d(2H $\rightarrow$ 2H, kernel=3, stride=1, pad=1) \\
Conv2d(2H $\rightarrow$ 2H, kernel=2, stride=2, pad=0) \\
BatchNorm(2H) ×3, LeakyReLU after each conv \\
Output shape: $(2H, 128, 128)$ \\ \hline

\textbf{Block 1:} \\
Conv2d(2H $\rightarrow$ 4H, kernel=3, stride=1, pad=1) \\
Conv2d(4H $\rightarrow$ 4H, kernel=3, stride=1, pad=1) \\
Conv2d(4H $\rightarrow$ 4H, kernel=2, stride=2, pad=0) \\
BatchNorm(4H) ×3, LeakyReLU ×3 \\
Output: $(4H, 64, 64)$ \\ \hline

\textbf{Block 2:} \\
Conv2d(4H $\rightarrow$ 8H, kernel=3, stride=1, pad=1) \\
Conv2d(8H $\rightarrow$ 8H, kernel=3, stride=1, pad=1) \\
Conv2d(8H $\rightarrow$ 8H, kernel=2, stride=2, pad=0) \\
BatchNorm(8H) ×3, LeakyReLU ×3 \\
Output: $(8H, 32, 32)$ \\ \hline

\textbf{Block 3:} \\
Conv2d(8H $\rightarrow$ 16H, kernel=3, stride=1, pad=1) \\
Conv2d(16H $\rightarrow$ 16H, kernel=3, stride=1, pad=1) \\
Conv2d(16H $\rightarrow$ 16H, kernel=2, stride=2, pad=0) \\
BatchNorm(16H) ×3, LeakyReLU ×3 \\
Output: $(16H, 16, 16)$ \\ \hline

\textbf{Block 4:} \\
Conv2d(16H $\rightarrow$ 32H, kernel=3, stride=1, pad=1) \\
Conv2d(32H $\rightarrow$ 32H, kernel=3, stride=1, pad=1) \\
Conv2d(32H $\rightarrow$ 32H, kernel=2, stride=2, pad=0) \\
BatchNorm(32H) ×3, LeakyReLU ×3 \\
Output: $(32H, 8, 8)$ \\ \hline

\textbf{Block 5:} \\
Conv2d(32H $\rightarrow$ 64H, kernel=3, stride=1, pad=1) \\
Conv2d(64H $\rightarrow$ 64H, kernel=3, stride=1, pad=1) \\
Conv2d(64H $\rightarrow$ 64H, kernel=2, stride=2, pad=0) \\
BatchNorm(64H) ×3, LeakyReLU ×3 \\
Output: $(64H, 4, 4)$ \\ \hline

\textbf{Bottleneck:} \\
Conv2d(64H $\rightarrow$ 128H, kernel=4, stride=1, pad=0) \\
BatchNorm(128H), LeakyReLU \\
Output: $(128H, 1, 1)$ \\ \hline

\textbf{Fully Connected:} \\
Flatten $\rightarrow$ Dropout \\
Linear(128H $\rightarrow$ 64H) + Dropout + LeakyReLU \\
Linear(64H $\rightarrow$ 12) \\ \hline

\textbf{Output:}
Last 6 values are squared to enforce positivity \\ \hline
\end{tabular}
\caption{\textbf{CNN architecture.} $H$ is the hidden dimension hyperparameter.}
\label{tab:arch}
\end{table*}



\end{document}